\documentstyle[aps,prl]{revtex}
\begin{document}
\draft
\twocolumn[
\title{Turbulence in Globally Coupled Maps}
\author{{\sc M. G. Cosenza and A. Parravano}}
\address{Centro de Astrof\'{\i}sica Te\'orica, Facultad de Ciencias, 
Universidad de Los Andes, \\A. Postal 26 La Hechicera, M\'erida 5251,
Venezuela} 
\date{To appear,  {\sf Phys. Rev. E, 1996}}
\maketitle
\begin{abstract}
\widetext
\vspace{-0.5cm}
The phenomenon of turbulence is investigated in the context of globally
coupled maps. The local dynamics is given by the Chat\'e-Manneville minimal 
map previously used in studies of spatiotemporal intermittency in
locally coupled map lattices. New features arise in the globally 
coupled system; for instance, the transition to turbulence takes place
discontinuously at some critical values of the parameters of the system.
The critical boundaries between different regimes (laminar, turbulent and
fully turbulent) of the system are calculated on the parameter space. 
Windows of turbulence are present on some ranges of the coupling parameter.
The system also exhibits nontrivial   
collective behavior.
A map for the instantaneous fraction of turbulent elements is proposed. 
This map describes many of the observed
properties of the system. 
\narrowtext
\end{abstract}
]

\section{Introduction.}
The transition to turbulence in confined systems and its relation to the
routes to chaos has been a subject of much interest \cite{Newell}.
A very general scenario for the occurrence of turbulence is spatiotemporal
intermittency, {\em i.e.}, a sustained regime characterized by the coexistence
of coherent-laminar and disordered-chaotic domains in space and time
\cite{Kan1,Exp}.
There are numerous studies of this phenomenon, some of the most
extensive of which have been on model dynamical systems such as coupled
map lattices (CML) \cite{Kan1,Chate,Houlrik,Bruyn,Grassberger,Co}.  
The idea is that the ingredients of a CML --- a discrete space, discrete time
system of interacting elements whose states vary continuously according to
specific functions --- are
sufficient to capture much of the phenomenology observed in complex 
spatiotemporal processes, in particular some relevant features of
spatiotemporal intermittency and turbulence. In this respect, a CML
can be viewed as a simplifying replacement for partial differential
equations of hydrodynamics \cite{Chate2}.
The onset of turbulence via spatiotemporal intermittency has mainly been
investigated on CMLs with local, diffusive-type interactions on both
low-dimensional Euclidean arrays \cite{Chate,Houlrik,Bruyn,Grassberger}
and on fractal lattices \cite{Co}. These studies have permitted the 
characterization of the transition to turbulence as a critical phenomenon
in one and two dimensions and on several fractal dimensions with diverse
local connectivities. 

Recently, globally coupled elements have been a focus of attention in
physics and biology \cite{Hadley,Roy,Marcus,Kuramoto,Kan2}, including
systems such as Josephson junction arrays, charge density waves,
multimode lasers, neural dynamics, ecological and evolution models.
Globally coupled maps \cite{KanGCM} constitute a useful approach
to the study of many processes on this kind of systems.
Globally coupled maps can be regarded as an extension of a CML with
diffusive coupling to infinite dimension, or as a limiting case of a system
with long range interactions. Spatial concepts lose meaning 
and only temporal properties become relevant
in  globally
coupled maps. These characteristics allow for simpler, mean-field 
descriptions of the behavior of the system.
  
In this paper, we investigate the phenomenon of 
turbulence in the context of globally
coupled maps.
The local dynamics we employ is the elementary map of Chat\'e and Manneville
\cite{Chate}, which has been shown to exhibit several properties 
of the transition to turbulence via spatiotemporal intermittency on
locally coupled map lattices \cite{Chate,Co}.
Our system of globally coupled Chat\'e and Manneville maps
provides a situation to compare the roles  that local and
global interactions play on the
occurrence of turbulence. Some new features arise in this model, for instance, 
the onset of turbulence appears to take place {\it discontinuously} at critical
values of the parameters of the system. 
The model is presented and its statistical properties are
numerically explored on its parameter space in Section~II. 
In section~III, a map for the instantaneous fraction of turbulent 
elements in the system is proposed. This map model describes much of 
the behavior observed in section~II.
The results are discussed in Section~IV.

\section{Globally coupled minimal maps for turbulence.}
We consider the globally coupled map system
\begin{equation}
\label{cml}
x_{t+1}(i)=(1-\epsilon)f(x_t(i))+\frac{\epsilon}{N}\sum_{j=1}^{N}f(x_t(j)),
\end{equation}
where $x_t(i)$ gives the state of the lattice element $i$ 
$(i=1,\ldots,,N=\mbox{system size})$ at 
a discrete time step $t$, 
$\epsilon$ is the coupling parameter, 
and $f$ is a map describing the local dynamics.
The instantaneous mean field of the system is defined as
\begin{equation}
\label{mf}
h_t=\frac{1}{N}\sum_{j=1}^{N}f(x_t(j)).
\end{equation}

While the choice of the local map $f$ in Eq.~(\ref{cml})
determines the details of the transition to turbulence, it need 
only possess a few general
properties to observe spatiotemporal intermittency. The map introduced
by Chat\'e and Manneville \cite{Chate} contains such minimal requirements,
\begin{equation}
\label{map} 
f(x)=\left\{
\begin{array}{ll}
\frac{r}{2}\left(1-\left| 1-2x \right|\right), & \mbox{if $x \in [0,1]$} \\
x, & \mbox{if $x > 1$},
\end{array}
\right.
\end{equation} 
with $r>2$. This map is chaotic as long as $f(x)$ remains in the interval 
$[0,1]$; however, when $f(x) > 1$ the dynamics reaches 
a fixed point in one iteration. The local variable can thus be seen as a 
continuum
of stable ``laminar" fixed points $(x>1)$ adjacent to a chaotic repeller
or ``turbulent" phase $(x \in [0,1])$. For a coupled system the laminar
phase is metastable, since sites in this state are stable to infinitesimal
perturbations but possibly unstable to finite disturbances induced by the
coupling.

In diffusively coupled map lattices, for a large enough coupling, the turbulent
phase can propagate through the lattice in time producing sustained states
of spatiotemporal intermittency \cite{Chate,Co}. Here, we investigate the
corresponding phenomenon in the context of globally coupled maps 
[Eq.~(\ref{cml})] using the local map $f$ [Eq.~(\ref{map})]. This situation
may be considered as spatiotemporal intermittency occurring on a 
diffusively coupled system in a spatially high dimensional lattice,
or as spatiotemporal intermittency in a limiting case of long range
interactions. Thus, the notions of spatial patterns and correlations 
characteristic of spatiotemporal intermittency in CMLs with local
connections are lost and only 
temporal properties of the system become relevant.
Above some threshold in parameter space 
$(r,\epsilon)$, and starting from random initial conditions, 
the globally coupled
map system  relaxes to a statistically stationary chaotic regime 
where each element displays intermittency between 
laminar and turbulent phases. As on locally coupled map lattices,
the transition to this
extended chaotic state can be monitored by the mean fraction of turbulent
sites $\langle F \rangle$, a quantity that acts as the order parameter for
the system \cite{Chate}. We have calculated  $\langle F \rangle$ as a function
of the coupling parameter $\epsilon$ for several fixed values of $r$ from
a time average of the instantaneous turbulent fraction $F_t$, as
\begin{equation}
\langle F \rangle={1 \over \tau} \sum_{t=1}^\tau F_t.
\label{av}
\end{equation}
Typically $10^4$ iterations were discarded before taking the
time average in Eq.~(\ref{av}) and $\tau$ was fixed at the value 
$10^4$. The typical system size used in the calculations
was $N=10^4$. We have found that increasing the lattice sizes does not 
appreciably
affect the results presented in this paper. The characteristics of the system
also persist for small lattice sizes (up to $N=100$).   
As initial conditions,
we use random cell values equally distributed between the
turbulent and the laminar ranges. It is to be noticed that a minimum 
number of initially turbulent cells is always required to reach
a sustained state of turbulence. 

Figures 1(a) and 1(b) present plots of $\langle F \rangle$ versus $\epsilon$
for two different fixed values of $r$. For small coupling, the system reaches
a uniformly laminar state. The onset of turbulence takes place discontinuously
at a critical
value of the coupling parameter, $\epsilon_c$. The mean turbulent fraction
$\langle F \rangle$ vanishes at a larger value of the coupling, giving raise
to a relaminarization of the system; {\it i.e.}, a second transition 
from turbulence to a uniform laminar state takes place at a value 
$\epsilon'_c > \epsilon_c$, establishing a well defined window 
of turbulence
on the range $[\epsilon_c,\epsilon'_c]$. Figure~1(b) shows than more than
one of such windows of turbulence, separated by laminar gaps, can occur in
globally coupled maps as the coupling is varied.

Windows of spatiotemporal intermittency on a range of the coupling parameter
have also been observed in coupled maps on fractal lattices with large
enough connectivities \cite{Co}. In general,
for CMLs with local
interactions, the transition from a laminar regime to spatiotemporal 
intermittency (and viceversa, in the case of windows of turbulence) 
behaves in many aspects as a second order phase transition, characterized
by the scaling relation 
$\langle F \rangle \sim (\epsilon-\epsilon_c)^\beta$, 
where $\beta$ is a critical exponent \cite{Chate,Houlrik,Co}. 
In contrast, the transition between laminar states and turbulence
in globally coupled maps 
appears as a discontinuous jump in the quantity $\langle F \rangle$,
a feature associated to first order phase transitions.
The absence of spatial relations in globally coupled maps rules out 
the possibility of supporting small
spatial domains of turbulent cells which would be necessary for
a continuous 
transition to turbulence.    

The error bars shown on the mean turbulent fraction $\langle F \rangle$
in figs.~1(a) and 1(b) correspond to the standard deviation (square root
of the variance)
of the time series of $F_t$ at each value of $\epsilon$. 
With increasing system size $N$, these fluctuations turn out
to decrease only up to some size, beyond which they remain constant. This 
phenomenon is  associated to a nontrivial 
collective behavior commonly observed
in globally coupled maps and which has been called ``violation of the law
of large numbers'' \cite{Kan}: the variance of the temporal
fluctuations of the mean field [Eq.~(\ref{mf})] does not scale as $N^{-1}$
for large $N$, but it saturates at some constant value. 
In the present case, the large amplitudes of the standard deviations
observed in figure~1 reflect collective periodic states of
the system. For example, figure~2 shows the return map of 
the instantaneous turbulent
fraction, $F_{t+1}$ vs. $F_t$ for parameter values $r=3$ and 
$\epsilon=0.295$, after discarding the transients. The instantaneous
turbulent fraction 
displays an approximate periodic behavior, with  
period {\it six}. Other nontrivial collective states can be observed
at different parameter values of the system.  
The instantaneous turbulent fraction  is a simpler 
statistical description than the mean field and, in our case, it already
manifests a collective periodic behavior of the system over long times, 
as shown 
in figure~2. Calculations of return maps of the mean field of the system
at parameter values giving turbulence should reveal
more details on the nature 
of those collective behaviors, such as 
global periodic attractors and 
possible quasiperiodic motions.

We have calculated numerically the critical values $\epsilon_c$, $\epsilon'_c$
corresponding to boundaries of the windows of turbulence observed in the system
as a function of $r$. There exists a maximum value $r_{\max}\simeq
3.29$ beyond which no turbulent regime is observed.  
In this case, the coupling cannot compensate the escape rate from the 
interval $[0,1]$ in the local maps.
Both parameters $r$ and $\epsilon$
were varied in $10^{-3}$ in order to detect the regions of turbulence. 
Figure~3 shows the complex critical boundary for the
onset of turbulence determined in this way. 
The structure of the turbulent windows
in parameter space
is revealed in fig.~3.
The fine structure of this 
critical boundary is actually more complex; there are several narrow
windows of turbulence and laminar gaps at smaller scales. 
Some of the complexity of the critical boundary between turbulent and
laminar regimes has already been suggested by less detailed calculations on
diffusively coupled one-dimensional lattices \cite{Houlrik,Bruyn}. 
For parameter values inside the critical boundary, the 
globally coupled map system exhibits sustained turbulence 
($0<\langle F \rangle \leq 1$). The region where full turbulence takes place 
$(\langle F \rangle=1)$ is also shown in Fig.~3.  
In this fully turbulent region, the globally coupled system, Eq.~(\ref{cml}),
uses only the regime $x \in [0,1]$ of the local map, Eq.~(\ref{map}).
Thus, the local dynamics is effectively a tent map with slope $r>2$.
The global coupling is capable of confining all the elements $x_t(i)$ to
the interval $[0,1]$, even though the local dynamics is repelling in this case.
As found by Kaneko \cite{Kan3}, a system of globally coupled tent maps with
$r<2$ presents collective behaviors (one- and two-band global
attractors) and a turbulent phase on different regions of its parameter plane 
$(r,\epsilon)$. One may expect that a simple extrapolation of 
the phase diagram in \cite{Kan3} should carry through this structure 
to the range $r>2$, 
corresponding to the fully turbulent region in Fig.~3.  
The laminar 
regime of the system 
($\langle F \rangle=0$) occurs for values $(r,\epsilon)$ outside the
critical boundary. This boundary signals the discontinuos transition
between the two collective states (laminar and turbulent) 
in the phase diagram of the system.  
Figure~3 also shows a theoretical critical boundary for the onset 
of turbulence 
in globally coupled maps 
as well as a predicted boundary for full turbulence, both 
obtained from a model presented in the next section. 

\section{Map for the Turbulent Fraction.}
The existence of two distinct states (laminar and turbulent)
in the local dynamics and the global coupling are simplifying features
that permit the construction
of a map for the instantaneous turbulent fraction in the system under
consideration.  
In this case, it is possible to estimate the instantaneous exchange
of cell
values 
between the two 
states in order to compute the 
change of the $F_t$ in one iteration. 

Let $\phi_t^{+}$ ($\phi_t^{-}$) be the fraction of cells that being laminar
(turbulent) at time $t$ will become turbulent (laminar) at time $t+1$.
Then, the change in the instantaneous turbulent fraction from one iteration
to the next is
\begin{equation}
\label{g}
\Delta F_t \equiv F_{t+1}-F_t=\phi_t^{+}-\phi_t^{-}.
\end{equation}

A cell $i$ that is laminar at time $t$ becomes turbulent at time $t+1$ if
\begin{equation}
(1-\epsilon )\,x_t(i)+\epsilon \,h_t<1,
\end{equation}
therefore,
$\phi_t^{+}$ is the fraction of cells satisfying the condition
$1\leq x_t(i)<x_m$, where
\begin{equation}
\label{xm}
x_m=\frac{1-\epsilon h_t}{1-\epsilon }.
\end{equation}
The quantity $x_m$ is the maximum value that laminar cells
may have in order to become turbulent in the next iteration.
Thus, the fraction of the laminar
range $[1,r/2]$ occupied by
cells making the transition to turbulence in the next iteration is
$(x_m-1)/(r/2-1)$.
Notice that
\begin{equation}
\label{eq0}
\begin{array}{lll}
\phi_t^{+}=1 , & {\mbox {if} } &\;  x_m>r/2, \\
\phi_t^{+}=0 , & {\mbox {if} } &\;  x_m<1 . \\
\end{array}
\end{equation}
For $x_m \in (1,r/2)$, we approximate the function $\phi_t^{+}$ as
\begin{equation}
\label{eq1}
\phi_t^{+}=
(1-F_t)\left(\frac{x_m-1}{r/2-1}\right)^{k_1},
\end{equation}
where the positive exponent $k_1$ is a parameter introduced to i
take into account the
non-homogeneous distribution of cells in the interval $[1,r/2]$.

Similarly, a cell $i$ with $x_t(i)\le 1/2$ becomes laminar
in the next
iteration if
\begin{equation}
(1-\epsilon )rx_t(i) +\epsilon h_t>1,
\end{equation}
while
the same transition for a cell
with $1/2<x_t(i)<1$ occurs if
\begin{equation}
(1-\epsilon )(1-x_t(i))r+\epsilon h_t>1,
\end{equation}
therefore,  $\phi_t^{-}$ corresponds to the fraction of cells
satisfying the condition $x_m/r<x_t(i)<1-x_m/r$.
Notice that
\begin{equation}
\begin{array}{lll}
\phi_t^{-}=0, & {\mbox {if} } &\;  x_m>r/2.
\end{array}
\end{equation}
For $x_m<r/2$, we assume the following form of the function $\phi_t^{-}$,
\begin{equation}
\phi_t^{-}=F_t\left(1-2\frac{x_m}{r}\right)^{k_2}.
\end{equation}
where the quantity ($1-2x_m/r$) is the fraction of the
turbulent range $[0,1]$  occupied by
cells that become laminar in the next iteration.
The positive exponent $k_2$ takes into account the
non-homogeneous distribution of cells in the interval $[0,1]$.

By using the above assumptions, the difference map Eq.~(\ref{g})
can be written as
\begin{equation}
\label{delta}
\Delta F_t= \left\{
\begin{array}{lll}
-F_t\left(1-\frac{2x_m}{r}\right)^{k_2}, & {\mbox {if} } &\; x_m\le 1 \\
(1-F_t) \left(\frac{x_m-1}{r/2-1}\right)^{k_1}
-F_t\left(1-\frac{2x_m}{r}\right)^{k_2}, & {\mbox {if} } &\; 1<x_m<r/2 \\
1-F_t , & {\mbox {if} } &\; x_m\ge r/2.
\end{array}
\right.
\end{equation}
The case $x_m\le 1$ (i.e $\phi_t^{+}=0$) describes
transient turbulent regimes evolving towards a final laminar state ($F_t=0$).
In the case $x_m\ge r/2$ (i.e $\phi_t^{-}=0$ and $\phi_t^{+}=1$),
the system reaches a fully turbulent regime ($F_t=1$) in one time step.

To obtain the explicit dependence of the
difference map Eq.~(\ref{delta}) on the parameters
$r$ and $\epsilon$, we need an expression for $h_t$.
The instantaneous mean field $h_t$ can be approximated as the sum of two
contributions
\begin{equation}
\label{mf2}
h_t=h_T(r,\epsilon)F_t+
h_L(r,\epsilon)\left(1-F_t\right),
\end{equation}
where $h_T$ and $h_L$ are the instantaneous mean fields of the
turbulent and of the laminar cells, respectively.
In particular, the instantaneous mean field $h_T$ of the turbulent cells
fluctuates for fixed values of $r$ and $\epsilon$.
When $h_T$ approaches the value $1$, the system tends to fall
to the laminar state; thus, the maximum values of $h_T$ in parameter
space are the relevant ones for the transition to turbulence.
From the numerical calculations, one can roughly approximate
$h_T$ and $h_L$ as:
\begin{eqnarray}
h_T\simeq & 0.48+1.05\epsilon, \\
h_L\simeq & 1+0.10(r-2).
\end{eqnarray}
The expression for $h_T$ reflects the fact that the
range of allowed values in the turbulent phase does not depend on $r$,
while an increase of the coupling $\epsilon$ enhances the
flow of laminar cells across the phase boundary $x=1$, producing
an increment of the population of turbulent cells with $x(i)>1/2$.
On the other hand,
since the range of allowed values in the laminar phase is $(r/2-1)$,
one could expect $h_L$ to increase with $r$.
The above approximations are sufficient
for the construction of a mapping of the turbulent fraction, Eq~(\ref{delta}),
that describes many relevant properties of the
globally coupled Chat\'e-Manneville maps.

Figure~4 shows $\Delta F_t$ vs. $F_t$ obtained from Eq~(\ref{delta})
for five
different values of $\epsilon$ and the fixed value $r=2.3$. The
exponents $k_1=0.8$ and $k_2=1.2$ were chosen in all the calculations since
they optimize the agreement of our model with the results from the
globally coupled map system, but the characteristics of the 
map model persist
for a range of values of the parameters $k_1$ and $k_2$.
The  real roots ($\Delta F_t=0$) of Eq~(\ref{delta}) give the fixed points
(stationary values) of the turbulent fraction $F_t$.
The laminar state $F_t=0$ is always a stable fixed point.
Additionally, two other fixed points, $F_1$ and $F_2$ $(F_1<F_2)$,
occur in the interval $\epsilon \in [\epsilon_c,\epsilon'_c]$,
as shown in Fig.~4.
The fixed point $F_1$ is unstable and it defines a critical value
of the instantaneous
turbulent fraction  bellow which the system invariably falls to the
the stable laminar state $F_t=0$. For values of the instantaneous
turbulent fraction above this critical value $F_1$,
the system evolves towards the upper fixed point $F_2$, which is stable.
As shown in figure~4 and in figure~5, the two non-vanishing 
fixed points coincide
at the critical values of the coupling, $\epsilon_c$ and $\epsilon'_c$,
corresponding to tangent bifurcations of the map Eq.~(\ref{delta}).
In figure~5 the fixed points of 
the map Eq.~(\ref{delta}) are plotted as a function of
$\epsilon$, for fixed $r=2.3$. Figure~5 clearly shows that at the
critical values of the coupling, $\epsilon_c$ and $\epsilon'_c$, the
transition between laminar $(F_t=0)$ and turbulent $(F_t=F_2)$
regimes predicted
by the map  Eq.~(\ref{delta}) is discontinuous, as seen in the
numerical calculations.
The mean turbulent fraction $\langle F \rangle$ 
obtained from direct simulations of
the globally coupled map system with $r=2.3$ is also shown as a
function of $\epsilon$ in fig.~5. The stable fixed point $F_2$ obtained
from the map model agrees well with the direct calculation.

The values $\epsilon_c$ and $\epsilon'_c$ corresponding to each value of
$r$ define the boundary between laminar and turbulent regimes
in the parameter space. The curves $\epsilon_c(r)$ and $\epsilon'_c(r)$
obtained from the model are shown in figure~2.
By comparing with the critical boundary resulting from direct calculations,
one can see that the model given by  Eq.~(\ref{delta}) provides a
qualitative description of the transition to turbulence in the globally
coupled map system.
For values of $r$ close to $2$ and small $\epsilon$,
the model reproduces quite well the
critical boundary.
The map model also predicts the existence of a boundary for the fully
turbulent region in the parameter plane. On this boundary,
$x_m=r/2$ and $h_t=h_T$; thus from Eq.~(\ref{xm}), we get
\begin{equation}
\label{fully}
r\simeq\frac{2}{1-\epsilon}\left[1-\epsilon(0.48+1.05\epsilon)\right].
\end{equation}
This curve is also plotted in fig.~2.
However, the simple model presented here fails to
reproduce the complex structure
of the critical boundary and it predicts a lower value for the
maximum value of $r$ for which turbulence can be sustained.

\section{Conclusions.}
Some of the features associated to the
occurrence of turbulence in globally coupled systems have been explored
in this paper. We have found that
the onset of turbulence occurs discontinuously at critical values
of the parameters, as in first order phase transitions. Previous studies
of diffusively coupled map lattices with Chat\'e-Manneville local
dynamics have shown that the transition to turbulence in those cases
is similar to a second order phase transition. 

We have made
detailed calculations that reveal the complexity of
the critical boundaries separating 
the different regimes of the system.
Multiple windows of turbulence and relaminarization of the system 
have been observed
as the coupling
parameter is varied.

Nontrivial collective behavior arises within the turbulent region in the 
parameter space of the globally coupled system. In fact, the return map 
of the instantaneous fraction of turbulent cells, which is a  
simple two-state statistical description of the system, 
reflects this collective temporal organization.   
The appearance of nontrivial collective behaviors is map lattices with
local or global couplings is a subject of current research. Much effort
has been devoted to establishing the necessary conditions for the emergence
of this behavior. 
Periodic collective states have been found in globally
coupled maps belonging to some universality class (tent, quadratic, or
circle maps) \cite{Kan3,Kan4}. 
The observation of collective periodic behavior  
in the present system, where the local dynamics has very
specific characteristics,
suggests that this kind of collective behavior should be a rather common 
phenomenon 
in deterministic systems of coupled chaotic elements.   

The proposed map for the instantaneous turbulent fraction explains
many of the observed aspects of the globally coupled Chat\'e-Manneville maps, 
such as:
the existence of the critical values of the coupling, $\epsilon_c$ and
$\epsilon'_c$, and the critical boundary in parameter space 
for the transition to turbulence; 
the discontinuos character of the transition;  
the existence of an $r_{\max}$; 
the presence of a threshold for the initial 
turbulent fraction in order to 
develop sustained turbulence; and 
the existence of a fully turbulent regime and its boundary in the parameter
plane.

The global map cannot describe the complex structure of the critical boundary
and it predicts a lower value of $r_{\max}$. The theoretical critical
boundary becomes  
less accurate, with respect to the direct simulations, for large
values of $r$ and $\epsilon$.
However, map models for global quantities, such as the mean field
or other statistical properties of the system,  
constitute a useful approach to the understanding  
of the collective dynamics of coupled map lattices.  
In particular, global maps analogous to  Eq.~(\ref{delta}) 
could model the behavior    
of other globally coupled systems whose local dynamics contains two distinct
states, as in bistable maps. 

\section*{Acknowledgments}
This work was supported by  grants C-717-95 and C-658-94
from the Consejo de Desarrollo
Cient\'{\i}fico, Human\'{\i}stico y Tecnol\'ogico of Universidad
de Los Andes, Venezuela. We thank the International 
Center for Theoretical Physics at Trieste for the hospitality while 
part of this work was carried out.

\begin{figure}
\caption{Mean turbulent fraction $\langle F \rangle$ 
as a function of the coupling parameter
$\epsilon$. The local parameter is fixed at
{\sf (a)} $r=3$; {\sf (b)} $r=2.6$. The error bars indicate 
$\pm 1$ standard deviations.} 
\end{figure}

\begin{figure}
\caption{Return map $F_{t+1}$ vs $F_t$ of the instantaneous turbulent fraction
obtained from direct simulation of the globally coupled Chat\'e-Manneville
maps. Parameters are $r=3$, $\epsilon=0.295$; and system size is $N=10^5$.
The number of iterations shown is $10^4$.}
\end{figure}

\begin{figure}
\caption{Phase diagram of the globally coupled map system.
The different stationary regimes of the system are indicated on the parameter
plane. The numerically determined critical boundaries between laminar and
turbulent states and between turbulent and fully turbulent regimes are
shown with thick line. The smooth (thin) curves correspond to the
theoretical prediction of those same boundaries; upper curve: theoretical 
laminar-turbulent boundary, lower curve: theoretical turbulent-fully turbulent
boundary (Section III).}     
\end{figure}

\begin{figure}
\caption{ 
The map $\Delta F_t$ vs. $F_t$ obtained from Eq~(14)
plotted for five labeled 
values of $\epsilon$ and fixed value $r=2.3$.
The critical values for the onset of turbulence are 
$\epsilon_c=0.0145$ and $\epsilon'_c=0.488$.}
\end{figure}

\begin{figure}
\caption{Numerically calculated mean turbulent fraction
$\langle F \rangle$ as a function of the
coupling $\epsilon$ for fixed $r=2.3$. The system becomes fully turbulent
($\langle F \rangle=1$) between $\epsilon \simeq 0.31$
and $\epsilon \simeq 0.39$.
The error bars indicate $\pm 1$ standard deviations.
The stable fixed point $F_2$ (thick line) and the unstable fixed point
$F_1$ (thin line) of the map Eq.~(14) are also shown as
functions of $\epsilon$ for $r=2.3$.}
\end{figure}

\end{document}